\documentclass[3p,times]{elsarticle}

\usepackage{ecrc}

\volume{00}

\firstpage{1}

\journalname{Nuclear Physics A}

\runauth{Jan Kapit\'an et al.}

\jid{nupha}

\usepackage{amssymb}

\usepackage[figuresright]{rotating}

\newcommand{\pT}{p_\mathrm{T}}
\newcommand{\kt}{k_\mathrm{T}}
\newcommand{\et}{E_\mathrm{T}}
\newcommand{\gev}{\mathrm{GeV}}
\newcommand{\gevc}{\mathrm{GeV}/c}

\begin{document}

\begin{frontmatter}



\dochead{}

\title{Jet studies in 200 GeV p+p and d+Au collisions from the STAR experiment at RHIC}


\author{Jan Kapit\'an (for the STAR Collaboration)}

\address{Nuclear Physics Institute ASCR, Na Truhlarce 39/64\\
18086 Praha 8\\
Czech Republic\\
kapitan@rcf.rhic.bnl.gov}

\begin{abstract}
Recent progress in full jet reconstruction in heavy-ion collisions at RHIC makes it a promising tool for the quantitative study of the QCD at high energy density. Measurements in d+Au collisions are important to disentangle initial state nuclear effects from medium-induced $\kt$ broadening and jet quenching. Furthermore, comparison to measurements in p+p gives access to cold nuclear matter effects. Inclusive jet $\pT$ spectra and di-jet correlations ($\kt$) in 200~GeV p+p and d+Au collisions from the 2007-2008 RHIC run are presented.  
\end{abstract}

\begin{keyword}
RHIC \sep Intial state effects \sep Jets \sep Di-jet correlations


\end{keyword}

\end{frontmatter}


\section{Introduction}
\label{intro}
Jets are remnants of hard-scattered partons, which are the fundamental objects of pQCD. At RHIC, they can be used as a calibrated probe of the hot and dense matter created in heavy ion collisions~\cite{Mateusz-QM09,Elena-QM09}. To quantify the signals observed in heavy ion collisions in comparison to p+p collisions, it is necessary to disentangle cold nuclear matter (CNM) effects from final state effects by measurements in systems such as d+Au. The CNM effects can be described by partonic rescattering~\cite{Vitev-InitialStateBroadeningPLB562} and by modification of parton distribution functions~\cite{EPS08}. Measurements of the single particle $R_\mathrm{dAu}$ have shown modification for intermediate $\pT$ even at midrapidity~\cite{BRAHMS-RdAu}. It is important to increase the kinematic reach in $R_\mathrm{dAu}$ measurements and jets are a good tool for this, as they are not prone to fragmentation biases. 

\section{Jet reconstruction}
\label{jets}
This analysis is based on $\sqrt{s_\mathrm{NN}} = 200~\gev$ data from the STAR experiment, recorded during RHIC run 8 (2007-2008). The Beam Beam Counter detector, located in the Au nucleus fragmentation region, was used to select the 20\% highest multiplicity events in d+Au collisions. The Barrel Electromagnetic Calorimeter (BEMC) detector was used to measure the neutral component of jets, and the Time Projection Chamber (TPC) detector was used to measure the charged component of jets. A so-called 100\% hadronic correction was used to combine tracking and calorimeter measurement in the case of a TPC track pointing to a BEMC tower: track momentum was subtracted from the tower energy to avoid energy double counting (electrons, MIP and possible hadronic showers in the BEMC). 

To minimize the biases due to possible BEMC backgrounds and dead areas, the jet neutral energy fraction is required to be within $(0.1,0.9)$. An upper $\pT < 15~\gevc$ cut was applied to TPC tracks due to uncertainties in TPC tracking performance at high-$\pT$ (under further investigation). The acceptance of TPC and BEMC together with experimental details (calibration, primary vertex position cuts) limit the jet fiducial acceptance to $|\eta|<0.55 \; (R=0.4), |\eta|<0.4 \; (R=0.5)$, where $R$ is the resolution parameter used in jet finding.

Recombination jet algorithms kt and anti-kt from the FastJet package~\cite{FastJet,AntiKt} are used for jet reconstruction. To subtract the d+Au underlying event background, a method based on active jet areas~\cite{JetAreas,BgSub} is applied: $\pT^{Rec} = \pT^{Candidate} - \rho \cdot A$, with $\rho$ (obtained event-by-event) estimating the background density and $A$ being the active jet area (active area for anti-kt jets with $R=0.4$ and jet $\pT>5~\gevc$ has mean value $0.52$ and RMS $0.05$). Due to the asymmetry of the colliding d+Au system, the background is asymmetric in $\eta$. This dependence (averaged over the whole event sample) was parametrised with a linear function in $\eta$ (see Figure~\ref{fig:etadep_bg}) and included in the background subtraction procedure. 

Pythia 6.410 and GEANT detector simulations were used to correct for experimental effects. Jet reconstruction was run at MC hadron level (PyMC) and at detector level (PyGe). To study residual effects of the d+Au background (such as background fluctuations), a sample with added background (PyBg) was created by embedding Pythia events into 0-20\% highest multiplicity d+Au events (minimum bias online trigger). This embedding was done at the level of reconstructed TPC tracks and BEMC towers.

\section{Nuclear $\kt$ broadening}
\label{ktbroadening}
Di-jet azimuthal correlations in p+p and d+Au can be used to estimate nuclear $\kt$ broadening. To increase the di-jet yield, a BEMC high tower (HT) online trigger (one tower with $\et > 4.3~\gev$) was used for both p+p and d+Au data. A resolution parameter $R=0.5$ was used for jet finding and a cut on $\pT > 0.5~\gevc$ was applied for tracks and towers to reduce background. To select a clean di-jet sample the two highest energetic jets ($p_\mathrm{T,1} > p_\mathrm{T,2}$) in each event were used, with $p_\mathrm{T,2} > 10~\gevc$. 

\begin{figure}[htb]
\begin{minipage}[h]{0.28\textwidth}
\centering
\includegraphics[width=1.0\textwidth]{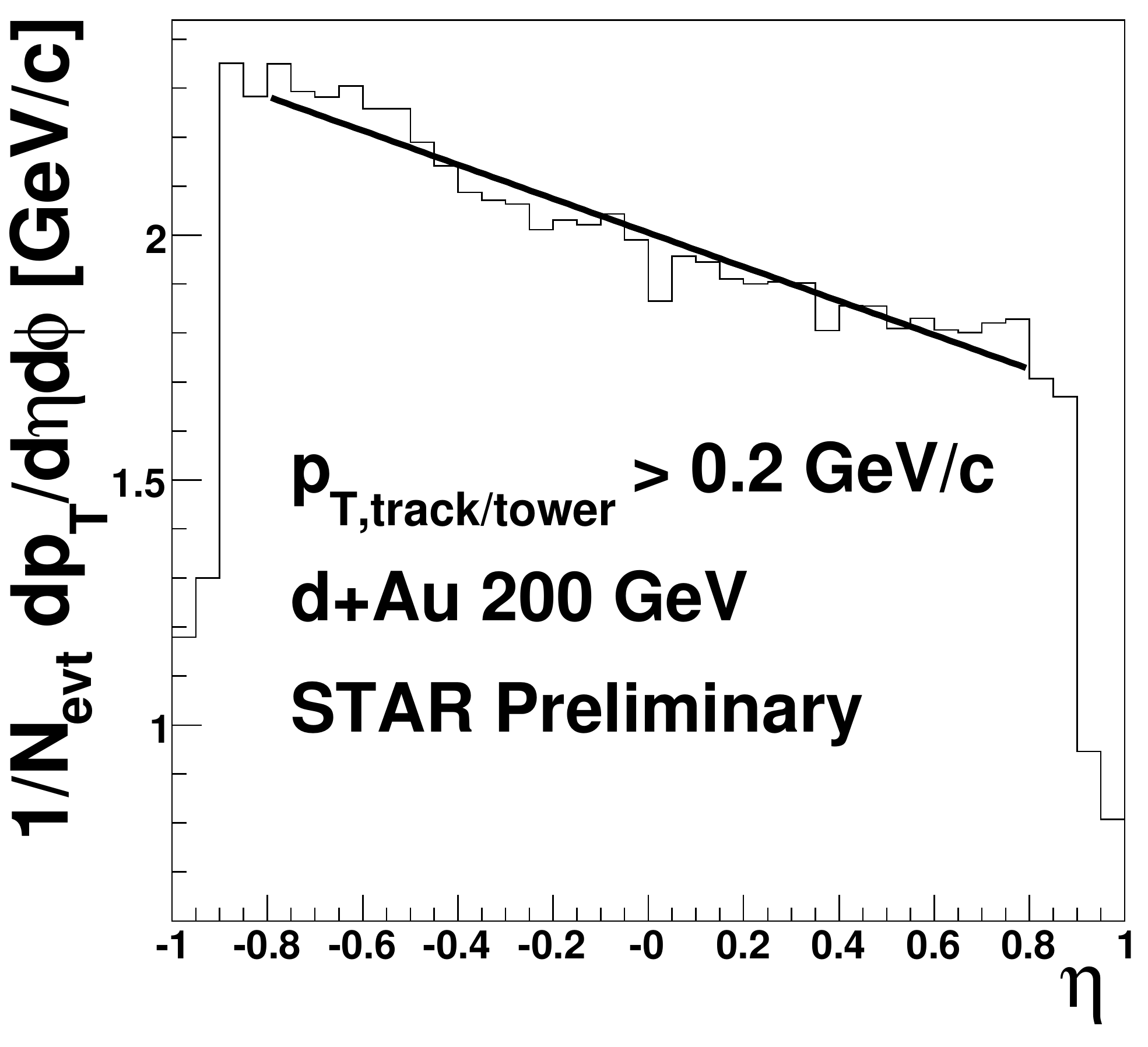}
\vspace{-0.6cm}
\caption{\label{fig:etadep_bg}Dependence of track + tower $\pT$ in minimum-bias triggered d+Au collisions on $\eta$, to determine the $\eta$ dependence of underlying event background.}
\end{minipage}
\hfill
\begin{minipage}[h]{0.71\textwidth}
\centering
\includegraphics[width=1.0\textwidth]{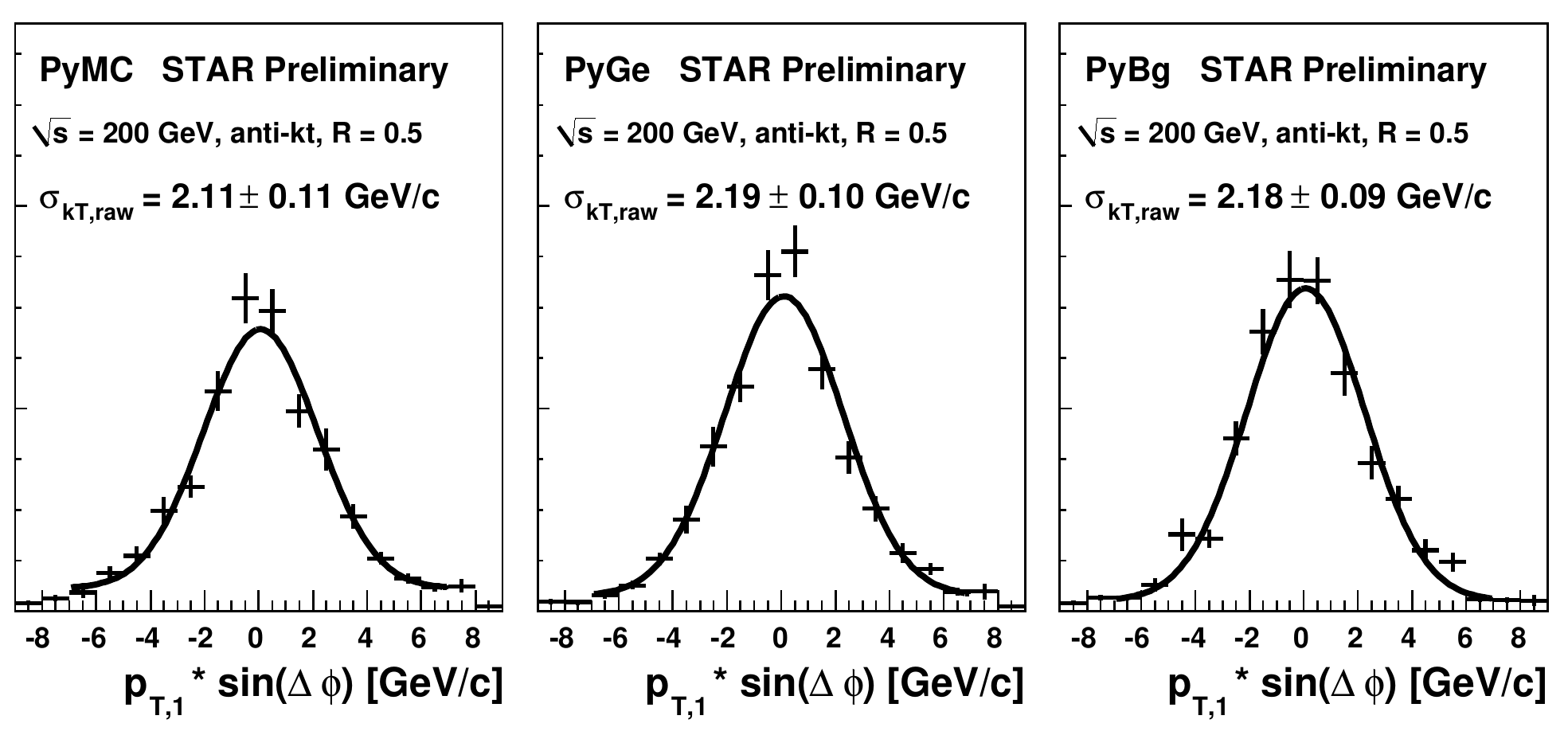}
\vspace{-0.85cm}
\caption{\label{fig:ktsimu}Distributions of $k_\mathrm{T,raw} = p_\mathrm{T,1} \sin(\Delta\phi)$ for simulation ($10 < p_\mathrm{T,2} < 20~\gevc$).}
\end{minipage}
\end{figure}

The total transverse momentum of a di-jet, $\kt$, is measured from its azimuthal component, experimentally accessible as $k_\mathrm{T,raw} = p_\mathrm{T,1} \sin(\Delta\phi)$, where $\Delta\phi$ is the di-jet opening angle. Gaussian widths, $\sigma_{k_\mathrm{T,raw}}$, were obtained for the two jet algorithms and two ($10 - 20~\gevc$, $20 - 30~\gevc$) $p_\mathrm{T,2}$ bins. Detector and residual background effects on the $\kt$ widths were studied by comparing PyMC, PyGe and PyBg distributions as shown in Figure~\ref{fig:ktsimu}. The widths are consistent within statistical errors, therefore detector and residual background effects can be approximately neglected.

Figure~\ref{fig:ktdata} shows the $k_\mathrm{T,raw}$ distributions in p+p and d+Au collisions for anti-kt algorithm and $10 < p_\mathrm{T,2} < 20~\gevc$. The Gaussian fit doesn't describe the p+p data very well and the precise shape of the distribution is currently under study. The RMS of these distributions have therefore been studied and they agree with the sigmas of the Gaussian fits.
The values extracted from the Gaussian fits are $\sigma_{k_\mathrm{T,raw}}^{p+p} = 2.8 \pm 0.1~\mathrm{(stat)}~\gevc$ and $\sigma_{k_\mathrm{T,raw}}^{d+Au} = 3.0 \pm 0.1~\mathrm{(stat)}~\gevc$. Therefore possible nuclear $\kt$ broadening seems rather small. 

\section{Inclusive jet spectra}
\label{spectra}
Run 8 d+Au data with a minimum bias online trigger were used to measure the jet $\pT$ spectrum. 10M 0-20\% highest multiplicity events after event cuts were used for jet finding (anti-kt algorithm) with a resolution parameter $R = 0.4$ and $\pT > 0.2~\gevc$ cut was applied to tracks and towers. The jet $\pT$ spectrum is normalized per event and the high multiplicity of d+Au events guarantees the trigger efficiency is independent of the $\pT$ of the hard scattering. Therefore, no correction related to trigger is applied.

Due to a high sensitivity of jet $\pT$ spectrum to the Jet Energy Scale (JES), an additional correction was applied to account for the lower TPC tracking efficiency in d+Au compared to that in the used p+p Pythia simulation. The d+Au tracking efficiency was determined by simulating single pions and embedding them at the raw detector level into real d+Au minimum bias events. The tracking efficiency in Pythia simulation was artificially lowered, prior to jet finding at PyGe and PyBg level, so that it matches the one obtained from d+Au embedding. 

\begin{figure}[htb]
\begin{minipage}[h]{0.6\textwidth}
\includegraphics[width=\textwidth]{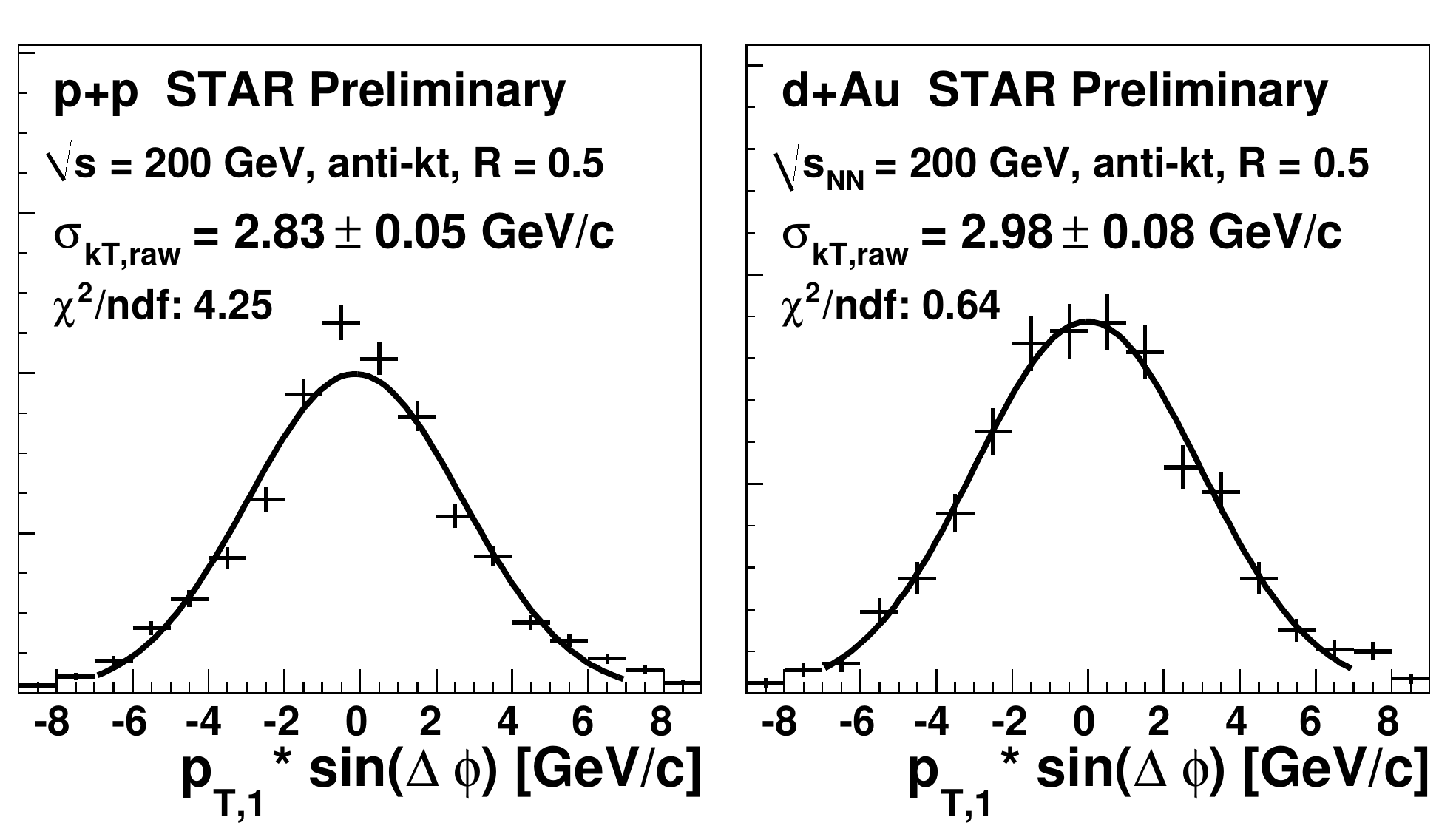}
\vspace{-0.75cm}
\caption{\label{fig:ktdata}Distributions of $k_\mathrm{T,raw}$ for p+p, d+Au ($10 < p_\mathrm{T,2} < 20~\gevc$).}
\end{minipage}
\hfill
\begin{minipage}[h]{0.38\textwidth}    
\includegraphics[width=\textwidth]{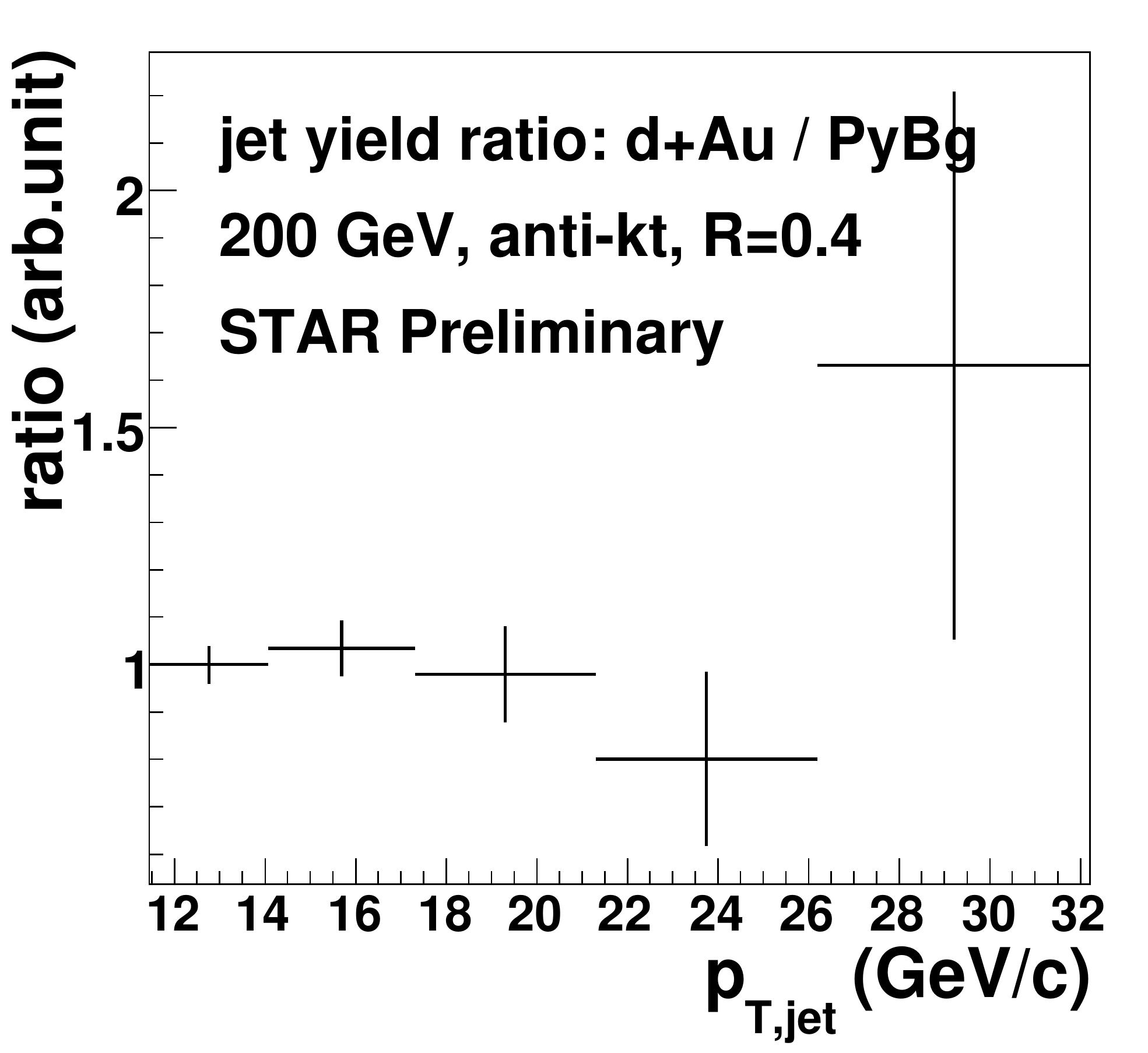}
\vspace{-0.84cm}
\caption{\label{fig:ratio}Ratio of jet $\pT$ spectra: d+Au/simulation.}
\end{minipage}
\end{figure}

A bin-by-bin correction is used to correct the jet spectrum to the hadron level. It is based on the generalized efficiency, constructed as the ratio of PyMC to PyBg jet $\pT$ spectra, applied to the measured jet $\pT$ spectrum. It therefore corrects for detector effects (tracking efficiency, unobserved neutral energy, jet $\pT$ resolution) as well as for residual background effects. As the impact of these effects on the jet $\pT$ spectrum differs substantially depending on the shape of the spectrum, the shapes have to be consistent between the PyBg and the measured jet $\pT$ spectra. Figure~\ref{fig:ratio} shows that this is indeed the case.

To compare the per event jet yield in d+Au to jet cross section measurements in p+p collisions, MC Glauber studies were utilized: $\langle N_\mathrm{bin} \rangle = 14.6 \pm 1.7$ for 0-20\% highest multiplicity d+Au collisions and $\sigma_\mathrm{inel,pp} = 42~\mathrm{mb}$. These factors were used to scale the p+p jet cross section measured previously by the STAR collaboration~\cite{STAR-ppJetPRL} using a Mid Point Cone (MPC) jet algorithm with a cone radius of $R = 0.4$.
The resulting d+Au jet $\pT$ spectrum is shown in Figure~\ref{fig:spectrum} together with the scaled p+p jet spectrum. The systematic errors are indicated by dashed lines and by the gray boxes. The dominant contribution to p+p systematic uncertainty is the JES uncertainty. Within these systematic uncertainties, the d+Au jet spectrum shows no significant deviation from the scaled p+p spectrum.

\section {Discussion on systematic uncertainties}
\label{systematics}
\begin{figure}[htb]
\centering
\includegraphics[width=0.74\textwidth]{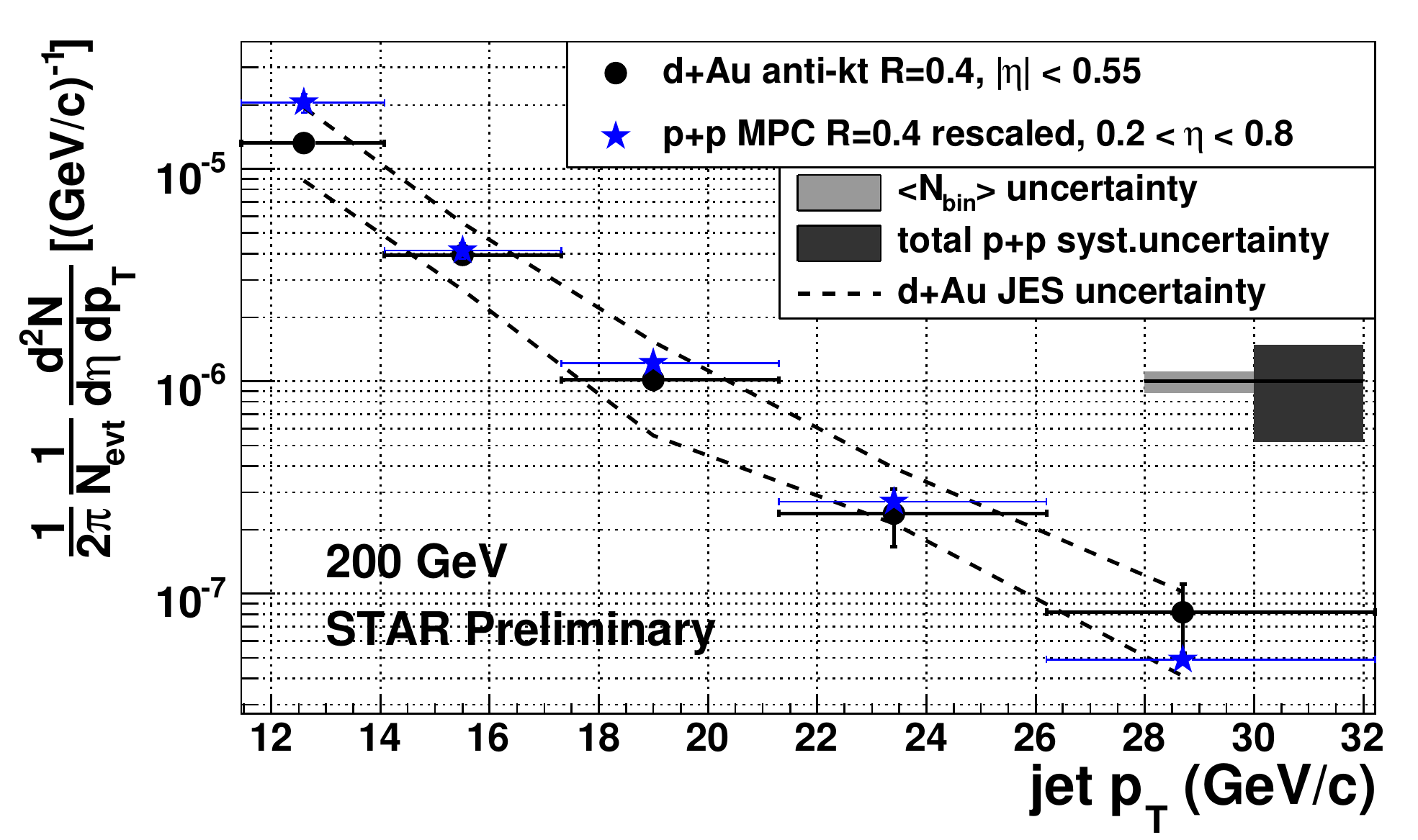}
\vspace{-0.35cm}
\caption{\label{fig:spectrum}Jet $\pT$ spectrum: d+Au collisions compared to scaled p+p~\protect\cite{STAR-ppJetPRL}.}
\end{figure}

The JES uncertainty dominates the uncertainties of d+Au measurement and is marked by the dashed lines in Figure~\ref{fig:spectrum}. Part of it comes from the BEMC calibration uncertainty of 5\%, applied to the neutral component of the jet. An uncertainty of 10\% in TPC tracking efficiency is applied to the charged component of jets. Embedding of jets into real d+Au events at raw detector level will allow to decrease this uncertainty in the future. As the JES uncertainty is expected to be largely correlated between run 8 p+p and d+Au data, we plan to measure jet $\pT$ spectrum in run 8 p+p collisions to decrease uncertainties in $R_\mathrm{dAu}$.

Caution is needed due to the use of different pseudorapidity acceptances and different jet algorithms in Figure~\ref{fig:spectrum}. Based on Pythia simulation, the effect of $\eta$ acceptance on jet $\pT$ spectrum is less than 10\% in the $\pT$ range covered by the present measurement. 
Various jet algorithms show different sensitivity to hadronization (important for low $\pT$ jets). Therefore, the same acceptance and the same algorithm should be used in p+p and d+Au to obtain the jet $R_\mathrm{dAu}$.

\section{Summary}
\label{summary}

Di-jet correlations and inclusive jet $\pT$ spectra were obtained from run 8 200 GeV p+p and d+Au collisions, using kt and anti-kt algoritms and the FastJet background subtraction scheme. 
No significant broadening due to Cold Nuclear Matter effects was observed in the measurement of di-jet $\kt$ widths.
In the studied kinematic range, the jet $\pT$ spectrum from 200 GeV 0-20\% highest multiplicity d+Au collisions is consistent with the scaled p+p jet spectrum within systematic uncertainties. These uncertainties have to be decreased in order to construct the jet $R_\mathrm{dAu}$ and this will be achieved by a precise tracking efficiency determination from jet embedding in raw d+Au data and jet cross section measurement in run 8 p+p data.

\section*{Acknowledgement}
\label{acknowledgement}

This work was supported in part by grants LC07048 and LA09013 of the Ministry of Education of the Czech Republic and by the grant SVV-2010-261 309.

\bibliographystyle{elsarticle-num}
\bibliography{HP2010proc_kapitan_v3}







\end{document}